\documentclass[aps,prb,preprint,groupedaddress,showpacs,showkeys]{revtex4}
\usepackage[T2A,OT1]{fontenc}
\usepackage[cp1251]{inputenc}         
\usepackage[russian,ukrainian,english]{babel} 
\usepackage[dvips]{epsfig}            
\usepackage{latexsym}
\usepackage{amsfonts}
\usepackage{boxedminipage}

\begin{document}

\title{Spin-polarized Current-induced Instability in Spin-Valve
with Antiferromagnetic Layer}
\author{ Helen V. Gomonay and Vadim M. Loktev}
\affiliation{National Technical University of Ukraine ``KPI''\\
37, ave Peremogy, Kyiv, 03056, Ukraine}

\begin{abstract}
In the framework of phenomenological model we consider dynamics of a
compensated collinear antiferromagnet (AFM) in the presence of
spin-polarised current. The model is based on the assumption that AFM spins
are localised and spin torque is transferred to each magnetic sublattice
independently. It is shown that in AFM spin current i) can be a source of
the "negative friction"; and ii) modifies spin-wave frequencies. Equilibrium
state of AFM can be destabilized by the current polarized in parallel to AFM
vector. Threshold current at which the loss of stability takes place depends
upon the magnetic anisotropy of AFM.
\end{abstract}
\keywords{antiferromagnet, spin-torque effect, spin-valve,
spin-polarised current} \maketitle

\section{Introduction}

The phenomenon of spin transfer from conductivity electrons to
magnetization of ferromagnetic (FM) layer is widely used in
engineering of the magnetic memory devices. While flowing from
nonmagnetic to ferromagnetic layer spin-polarised electrons
transfer spin torque \cite{Slonczewski:1996} and additional
magnetization \cite{Gulyaev:2006} thus inducing reorientation or
even dynamically stable rotation of localised magnetic moments.
Physical interpretation of these phenomena is based on the law of
spin conservation and $s-d$ exchange interaction between free
carriers and localized moments.\cite{Slonczewski:1996,
stiles:2004}

Recent experiments with nanopillars \cite{wei:2007, Urazhdin:2007}
point out that spin-polarised current also can change the state of
antiferromagnetic (AFM) layer and characteristic value of critical
current at which reorientation of spins takes place could be much
smaller than in FM. From general point of view, study of spin
transport effects in AFM may open either more efficient methods
for spintronics or much more reach fundamental phenomena. In
particular, in the AFM metal with spin-density waves (SDW) $s-d$
exchange couples spins of free electrons with orientation of AFM
vector and influence of spin torque is substantially
enhanced.\cite{Nunez:2005}

In the present paper we address another question: ``Is it possible to
control the state of an AFM metal \textbf{without} SDW with spin-polarized
current?'' As a starting point we consider the ``toy'' model of the
compensated collinear AFM in which the magnetic order is mainly caused by
localised spins. Fe$_{50}$Mn$_{50}$ alloy widely used in spin-valve
structures can be considered as an example of such a material.

\section{Model}

\begin{figure}[htbp]
{\centerline{\includegraphics[width=0.5\textwidth]{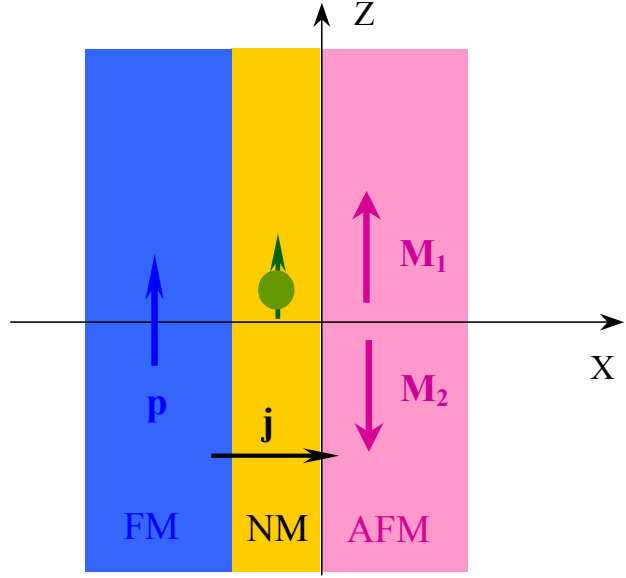}}
\caption{(Color online) Spin-valve structure consisting of FM and
AFM layers and nonmagnetic (NM) spacer. In the standard spin-valve
structure \cite{Urazhdin:2007} NM spacer is
absent.\label{fig_1}}}\end{figure}

We consider a spin-valve structure (analogous to that studied in
Ref. \onlinecite{wei:2007, Urazhdin:2007} consisting of FM and AFM
layers separated by a nonmagnetic metallic spacer
(Fig.\ref{fig_1})
 rather thin in order to
condition a ballistic regime for conductivity electrons. FM with
an easy axis in $\mathbf{p}$ direction acts as a spin polarizer
for the electron current $I$ flowing through the whole structure.
The magnetic state of AFM is unambiguously defined by sublattice
magnetizations $\mathbf{M}_{j}$ ($j$=1,2). We assume that due to a
local character of $s-d$ exchange, spin conservation law is
fulfilled independently for each act of conductivity-to-localized
spin interaction \cite{Haney:2007(2)}. So, according to
Slonczewskii mechanism \cite{Slonczewski:1996}, each sublattice
magnetization experiences a spin torque $\mathbf{T}_{j}=\sigma
I[\mathbf{M}_{j}\times[\mathbf{M}_{j}\times\mathbf{p}]]/M_{0}$,
where coefficient $\sigma =\varepsilon \eta \mu_{0}g/(2M_{0}Ve)$
depends upon the geometry of contact (volume $V$), and
spin-polarization efficiency $\varepsilon $, $M_{0}=\vert
\mathbf{M}_{j}\vert$. Here $\eta $ is the Plank constant, $g$ is
gyromagnetic ratio and $e$ is electron charge. Depending on the
direction of the electron current the sign of $I$ can be either
positive (incoming spin flux) or negative (outcoming spin flux).

 In the present paper we
restrict ourselves with the case of small external exposure, i.e.,
the work of the polarized current over the localized spins is
supposed to be much smaller than the exchange energy that keeps
magnetizations $\mathbf{M}_{1}$, $\mathbf{M}_{2}$ antiparallel. To
this end, macroscopic magnetization
$\mathbf{M}=\mathbf{M}_{1}+\mathbf{M}_{2}$ is much smaller than
AFM vector $\mathbf{L}=\mathbf{M}_{1}-\mathbf{M}_{2}$, $\vert
\mathbf{M}\vert \ll \vert \mathbf{L}\vert$, and one can reduce the
description of AFM dynamics to a Lagrange form with $\mathbf{L}$
as a generalized variable and the Lagrange function in a form
\cite{Bar:1986E}
\begin{equation}
\label{Lagrangian} L = \frac{\chi _ \bot }{8g^2M_0^2 }{\rm {\bf
\dot {L}}}^2 - \frac{A}{8M_0^2 }\left( {\nabla {\rm {\bf L}}}
\right)^2 - w_{an} ({\rm {\bf L}}).
\end{equation}

Here $\chi_\perp$, $A$, and $w_{an}(\mathbf{l})$ are the magnetic
susceptibility, inhomogeneous exchange constant and magnetic
energy of AFM layer, respectively.

Within the framework of the Lagrange formalism, all the dissipative
processes (Gilbert damping and spin-torque induced rotation of magnetic
moments) could be adequately described with the Relay function
\begin{equation}
\label{Relay} R = \frac{\chi _ \bot \alpha _G }{8g^2M_0^2 }{\rm
{\bf \dot {L}}}^2 - \frac{\sigma I}{4gM_0 }({\rm {\bf p}},{\rm
{\bf L}}\times {\rm {\bf \dot {L}}}),
\end{equation}
\noindent where the damping parameter $\alpha_{G}$ is equal to the
linewidth of AFM resonance (see, e.g.,
Ref.\onlinecite{Bar'yakhtar:1984}).

\section{General considerations}

Some peculiarities of AFM dynamics in the presence of
spin-polarized current can be deduced from the analysis of the
Relay function (\ref{Relay}) that describes the rate of energy
losses in the system.
\begin{itemize}
\item\textbf{Like} in FM, spin-polarized current may work as a
source of the external energy pumping (``negative'' friction) and
suppress the Gilbert damping. This takes place for a certain value
of current, $I~ \ge I_{c1}$, and noncollinear orientation of FM
and AFM easy axes, e.g., $\mathbf{p} \perp \mathbf{L}^{(0)}$.
Critical current $I_{cr}$ at which the effective damping changes
sign is calculated from the condition of negative dissipation ${
\dot \mathbf{L}}(\partial R / \partial {\dot \mathbf{L}}) \le 0$.
In the case of steady precession of AFM vector with a frequency
$\omega $ around the equilibrium direction $\mathbf{L}^{(0)}$, the
critical current is given by the expression $I_{\rm cr} \propto
\chi_{ \perp }\alpha_{G}\omega /\sigma $, analogous to that in FM
material.\cite{Slavin:2005} \textbf{In contrast to} FM, the value
of critical current in AFM is substantially reduced due to strong
exchange interaction between the magnetic sublattices (AFM
susceptibility $\chi_{\perp}$ is small).

\item An efficient energy pumping takes place for the precessional
motion only, i.e., when $\mathbf{L}\perp \dot \mathbf{L}$. Linear
oscillations of AFM vector (with $\mathbf{L}\|\dot \mathbf{L}$)
are always dissipative, $\dot \mathbf{L}(\partial R/\partial \dot
\mathbf{L})\ge 0$.

\item \textbf{Unlike} FM, the presence of spin-polarized current may
change spin-wave spectra of AFM and thus give rise to instability
and a kind of spin-flop transition in the case when FM and AFM
easy axes are parallel, $\mathbf{p}\| \mathbf{L}^{(0)}$. As can be
seen from (\ref{Relay}), small deviations $\delta
\mathbf{L}_{\perp}\perp\mathbf{L}^{(0)}$ oriented perpendicular to
equilibrium vector $\mathbf{ L}^{(0)}$ induce a generalized force
$\mathbf{F} = - \partial R / \partial  \dot{ \mathbf{L}}=
\mathbf{p}\times \delta \mathbf{L}_{\perp}(I\sigma /4{gM}_{0})$.
This force is a linear function of $\delta \mathbf{L}_{\perp}$ and
thus may compete with the restoring force produced by the magnetic
anisotropy field.
\end{itemize} In the next section we consider the last case in more
details.

\section{Current-induced instability }

The typical AFM metal used in spin-valves can be thought of as an
``easy-plane'' AFM because of \textit{i}) a very small anisotropy
of bulk materials and \textit{ii}) possible out-of-plane
anisotropy produced by the shape and interfacial interactions. Let
equilibrium orientation of AFM vector $\mathbf{L}^{(0)}$ be
parallel to FM magnetization $\mathbf{p}\|Z$ in the film plane. In
this case the linearized Lagrange equations for small excitations
$L_{x}$, $L_{y }$ are obtained from (\ref{Lagrangian}),
(\ref{Relay}) as follows:
\begin{eqnarray}
\label{eq3} \ddot {L}_x + \alpha _G \dot {L}_x - c^2\nabla ^2L_x +
\omega _x^2 (0)L_x - \sigma IgM_0 \chi _ \perp ^{- 1} L_y =
0,\nonumber\\
\ddot {L}_y + \alpha _G \dot {L}_y - c^2\nabla ^2L_y + \omega _y^2
(0)L_y + \sigma IgM_0 \chi _ \perp ^{ - 1} L_x = 0,
\end{eqnarray}
\noindent where the gaps $\omega _j (0) = g\sqrt {K_j / \chi _
\perp } $, ($j=x, y$) in spin-wave spectra are expressed through
the effective anisotropy constants $K_{x}$, $K_{y}$, $c = g\sqrt
{A / \chi _ \perp}$ is a spin-wave velocity, $X$ axis is directed
perpendicular to the film plane.

The analysis of shows that depending on the current value $I$ equations (\ref{eq3}),
 have two types of solutions. Below the threshold $I\le I_{\rm th1}
\equiv g\vert K_{x}-K_{y}\vert /(2M_{0}\sigma )$ AFM vector
oscillates around equilibrium direction with eigenfrequencies
\begin{equation}
\label{eq5} \Omega _{1,2}^2 (\mathbf k) = \frac{1}{2}\left(
{\omega _x^2 + \omega _y^2 } \right) + c^2{\mathbf k}^2\pm
\frac{1}{2}\left( {\omega _x^2 - \omega _y^2 } \right)\sqrt {1 -
(I / I_{\rm{th1}} )^2} ,
\end{equation}
\noindent where $\mathbf{k}$ is wave-vector. Both modes are
linearly polarized. The greater the current, the greater is the
out-of-plane component $L_{x}$. Energy dissipation is due to
internal friction solely and thus, equilibrium state with
$\mathbf{L}^{(0)}\| \mathbf{p}$ is stable.

With increase of $I$ the difference between the frequencies
$\Omega _{1 }$ and $\Omega _{2}$ decreases until at $I=I_{\rm th1
}$ the spectrum became degenerate, $\Omega _{1}=\Omega_{2}$.
Polarization of eigen modes can be either linear or circular and
energy dissipation is governed by two mechanisms: damping and
pumping. In the interval $I_{\rm{th1}} \le I \le I_{\rm{th2}}
\equiv \sqrt {I_{\rm{th1}}^{2} + I_{\rm{cr}}^{2}}$ damping is
stronger that pumping and the state with $\mathbf{L}^{(0)}\|
\mathbf{p}$ is stable. The value of critical current
\begin{equation}
\label{eq6} I_{\rm{cr}} \equiv \frac{\alpha_{G} \sqrt {K_y \chi _
\perp} }{2M_0 \sigma} = \frac{\alpha_{G}}{\omega_y (0)}\frac{K_y
}{K_x - K_y }I_{\rm th1}
\end{equation}

\noindent is calculated from the condition of accurate
compensation of two dissipation mechanisms. For definiteness we
assume that in-plane anisotropy $K_{y}$ is weaker than
out-of-plane $K_{x}$.

At $I\ge I_{\rm th2}$ an amplitude of at least one of the modes
grows exponentially with the current-dependent increment $\alpha =
\alpha_{G} (I - I_{\rm{th2}} ) / I_{\rm{cr}}$. This means that the
state with $\mathbf{L}^{(0)}\|\mathbf{p}$ becomes unstable and the
system evolves to a new state, e.g. to another (nonparallel to
$\mathbf{p}$) equilibrium orientation of AFM vector in the film
plane. Such a behaviour is somehow analogous to spin-flop
transition observed in AFMs of the ``easy-plane'' type under the
action of external magnetic field applied in parallel to AFM
vector.

\section{Discussion}

\begin{figure}[htbp]
{\centerline{\includegraphics[width=0.5\textwidth]{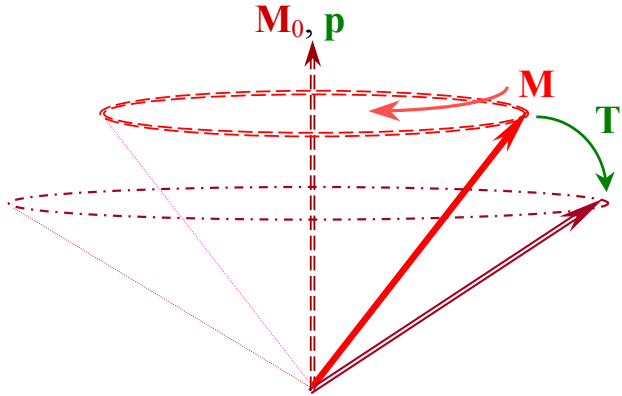}}
\caption{(Color online) Rotation of magnetization $\mathbf{M}$
under the action of spin torque.\label{fig_2}}}\end{figure}

\begin{figure}[htbp]
{\centerline{\includegraphics[width=0.5\textwidth]{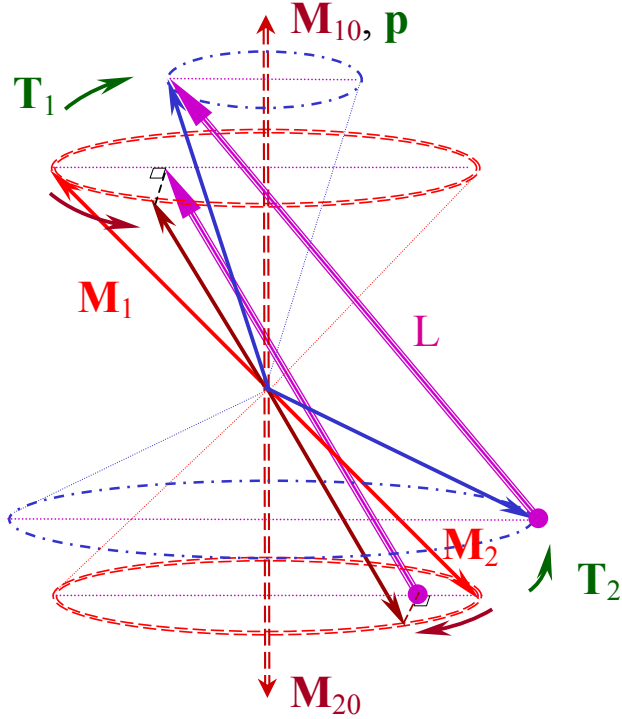}}
\caption{(Color online) Rotation of sublattice magnetizations
$\mathbf{M}_{1,2}$ and AFM vector $\mathbf{L}$ under the action of
spin torques $\mathbf{T}_{1,2}$.\label{fig_3}}}\end{figure}

The described dynamics of AFM in the presence of spin-polarised
current differs substantially from that in FM materials. The
difference can be intuitively understood from the geometry of spin
rotation (see Figs.\ref{fig_2} and \ref{fig_3}). In the FM
characterised by a single magnetic vector $\mathbf{M}$
magnetization has only two degrees of freedom. In the absence of
any dissipative processes magnetic excitations take a form of
precessional motion of $\mathbf{M}$ around its equilibrium
direction $\mathbf{M}_{0}$ (double-line ellipse in
Fig.\ref{fig_2}). The motive force of the precession is an
effective internal field that keeps magnetization direction along
an easy axis $\mathbf{M}_{0}$ (in the particular case, parallel to
spin polarisation axis $\mathbf{p}$). Spin torque $\mathbf{T}$
acts in such a way as to change an angle between $\mathbf{M}$ and
$\mathbf{M}_{0}$, and, correspondingly, energy of excitation.
Thus, in FM spin torque always acts as an energy source (or drain)
and thus its effect is equivalent to positive/negative friction.
Nondissipative dynamics in FM is possible only in the case of
precise balance between the torque-induced pumping and internal
damping.

AFM with two magnetic sublattices has more degrees of freedom. In
the absence of dissipation the low-energy excitations correspond
to coherent precession of both sublattice magnetizations
$\mathbf{M}_{1}$ and $\mathbf{M}_{2}$ (double-line ellipses in
Fig.\ref{fig_3}). The effective internal fields rotate
magnetizations in opposite directions so that an AFM $\mathbf{L}$
vector can oscillate within a plane.

Spin torques $\mathbf{T}_{1}$ and $\mathbf{T}_{2}$ turn both
magnetizations $\mathbf{M}_{1}$ and $\mathbf{M}_{2}$ in the same
direction (``up'' in Fig.\ref{fig_3}).

So, for one sublattice an angle between the excited and
equilibrium orientations increases and for another sublattice
decreases. This means that for a certain relation between spin
current and internal field (defined by the magnetic anisotropy
constants $K_{x,y})$ corresponding changes in energy could be
totally compensated and torque-induced motion is nondissipative
even in the absence of the internal damping.

The described (``nondissipative'') influence of spin-current on
AFM below the threshold current $I<I_{\rm th1}$ is to a certain
extent analogous to the affect of an external magnetic field
applied in parallel to $\mathbf{L}^{(0)}$. Both the magnetic field
and spin-current may give rise to the softening of one of the
spin-wave modes and cause spin-flop transition or transition to
dynamically stable stationary state. Each effect is insensitive to
the reversal of field direction, spin polarization and current
direction. On the contrary, due to the difference of symmetry
properties and depending on mutual orientation of field, spin and
current flow, combined application of the magnetic field and
spin-polarized current, may give rise to an enhancement or to
reduction of threshold current and spin-flop field. Detailed
analysis of this situation is beyond the scope of the paper.

The dynamics obtained in the framework of a very simple ``toy''
model is nevertheless qualitatively consistent with the observed
\cite{Urazhdin:2007} direct effect of electron current on the
magnetic state of FeMn, namely, irreversible switching of
spin-valve structure at a threshold current $I \propto 5\div
7.5$~mA. The effect was observed in the presence of the external
magnetic field $H \propto 0.1$ T. The authors attributed this
behavior to ``reorientations of magnetic configuration of FeMn
among a few metastable states''. We think that the reason of
reorientation can be the current-induced instability described
above. External field applied in-parallel to $\mathbf{L}^{(0)}$ is
a source of additional magnetic anisotropy.

The value of threshold current can be roughly estimated using the
value of FeMn bulk susceptibility \cite{Endoh:1971} $\chi_{\perp
 }=10^{- 5}$ (SI units), magnetization $2\mu _{0}M_{0}$=0.1 T
and typical AFM layer dimensions \cite{Urazhdin:2007} 120x60x1.5
nm$^{3}$. We assume that out-of-plane anisotropy $K_{x}$ can be as
large as 10$^{5 }$J/m$^{3}$ due to the interface effects (e.g.,
coupling strength between Co/FeMn layers is estimated
\cite{kuch:2002} as 10$^{-4}$~J/m$^{2}$ and monolayer thickness is
$ \propto 3 \cdot 10^{ - 10}$~m). Altimately, in the case of
100{\%} spin-polarization efficiency, we get $I_{th1} \propto
K_{x}Ve/\eta  \propto 10$~mA. Threshold current $I_{\rm th2}$,
which separates reversible/irreversible rotation of vector
$\mathbf{L}$ is of the same order of value, at least in the case
of the pronounced anisotropy ($K_{x}-K_{y} \propto K_{y})$ and
quality factor of AFM resonance $\omega_{y}/\alpha_{G} \ge 10$.
Really, in this case, as follows from (\ref{eq6}), critical
current $I_{\rm cr} \le 0.1 I_{\rm th1}$ and $I_{\rm th2} \approx
I_{\rm th1} \propto 10$~mA that agrees in order of value with
experimental results.

\section{Conclusions}

In summary, we have considered the dynamics of a compensated AFM
with localized spins in the presence of spin-polarized current. In
contrast to FM, spin current is not only a source of ``negative
friction'' but it also acts as an ``effective field'' that
modifies spin-wave modes and gives rise to the loss of stability
of the state with parallel orientation of spin polarization and
AFM vector. Predictions of the above ``toy'' model are in
qualitative agreement with experimentally observed influence of
spin current on AFM FeMn. Estimated values of the threshold
current are of the same order of value as characteristic currents
in the experiment.\cite{Urazhdin:2007}

\acknowledgements H.G. is grateful to Prof. A. N. Slavin for
discussions and drawing her attention to the problem considered in
the paper.


\end{document}